\documentclass{emulateapj}
 
\usepackage{amssymb,amsmath,bm}
\usepackage{graphicx}
\usepackage[colorlinks]{hyperref}
\usepackage{natbib}
\usepackage{graphicx}
\usepackage{subfigure}

\begin{document}

\title{Vector interaction enhanced bag model for astrophysical applications}

\author{Thomas~Kl{\"a}hn$^*$ and Tobias~Fischer}

\affil{Institute for Theoretical Physics, University of Wroc{\l}aw, pl. M. Borna 9, 50-204 Wroc{\l}aw, Poland\\
     $^*$ thomas.klaehn@ift.uni.wroc.pl}

\begin{abstract}
For quark matter studies in astrophysics the thermodynamic bag model (tdBAG) has been widely used. Despite its success it fails to account for various phenomena expected from Quantum-Chromo-Dynamics (QCD). We suggest a straightforward extension of tdBAG in order to take the dynamical breaking of chiral symmetry and 
the influence of vector interactions explicitly into account.
As for tdBAG the model mimics confinement in a phenomenological approach.
It is based on an analysis of the Nambu--Jona-Lasinio (NJL) model at finite density. 
Furthermore, we demonstrate how NJL and bag models in this regime follow from the more general and QCD based framework of Dyson-Schwinger (DS) equations in medium by assuming a simple gluon contact interaction. 
Based on our simple and novel model, we construct quark hadron hybrid equations of state (EoS) and study systematically chiral and deconfinement phase transitions, the appearance of $s$-quarks and the role of vector interaction. 
We further study these aspects for matter in $\beta$-equilibrium at zero temperature, with particular focus on the current $\sim2$~M$_{\odot}$ maximum mass constraint for neutron stars.
Our approach indicates that the currently only theoretical evidence for the hypothesis of stable strange matter is an artifact of tdBAG and results from neglecting the dynamical  breaking of chiral symmetry.
\end{abstract}

\date{\today}

\keywords{dense matter --- equation of state --- elementary particles: quarks --- stars: neutron}

\maketitle

\bibliographystyle{apj}

\section{introduction}

QCD is believed to be the correct theory of strongly interacting matter.
It inspires multifaceted research, both, theoretical and experimental (\citet{Brodsky:2015aia}
highlight this statement concerning the impact of QCD on hadron physics). 
Lattice QCD as the ab-initio approach to solve QCD numerically is increasingly successful in vacuum, at finite 
temperatures and at small chemical potentials~\citep[cf.][and references therein]{Fodor:2004,Aoki:2006}. 
It is not suited to provide solutions for low temperatures at chemical potentials in vicinity of the predicted deconfinement phase transition. 
This is the domain of interest for astrophysical applications, e.g., simulations of neutron stars (NS) and core-collapse
supernovae. 
The influence of physically realized deconfined quark matter in NS/protoneutron stars and supernovae 
on potential observables is a long standing research topic
~\citep[cf.][]{Alford:2006vz,Fischer:2010wp,Klahn:2006ir,Klahn:2006iw,Klahn:2011fb,Nakazato:2008su,Pagliara:2009dg,Pons:2001ar,Sagert:2008ka,Schertler:2000xq}.

As an alternative to lattice calculations the DS approach solves QCD's gap equations 
within a given truncation scheme. 
It is applicable in the whole temperature--density domain of the QCD phase diagram. 
Although the DS formalism has been successfully applied to understand 
hadron physics on the quark level \citep[cf.][and references therein]{Bashir:2012fs,Cloet:2013jya,Chang:2011vu,Roberts:2012sv} 
only little work has been done to exploit it
at finite densities \citep[cf.][]{Roberts:2000aa} and to describe the EoS of deconfined quark matter  \citep[cf.][]{Chen:2008zr,Chen:2011my,Chen:2015mda,Klahn:2009mb}.
These explorative studies promise deeper
insights into the EoS of cold, dense matter in a domain characterized by non-perturbative QCD.

Perturbative QCD, valid near the limit of asymptotic freedom where quarks are no longer strongly
coupled provides a valuable benchmark for DS in-medium studies. 
Recent work by  \citet{Kurkela:2014vha} illustrates how perpurbative QCD 
can be used in a reasonable way to put asymptotic high-density constraints on the EoS for compact stars.

The majority of studies of dense quark matter in astrophysical systems is based on effective models. 
These mimic certain but not necessarily all key features inherent to QCD, e.g., the breaking/restoration 
of chiral symmetry and the effect of deconfinement on the EoS and related quantities. 

In this work we demonstrate that the two most commonly used effective models in astrophysics, 
namely tdBAG as introduced in \citet{Farhi:1984qu}, 
and models of the NJL type \citep[cf.][]{Nambu:1961tp,Klevansky:1992qe,Buballa:2003qv} 
can both be understood as solutions of QCD's in-medium DS gap equations 
within a particular set of approximations. 
From this perspective we suggest modifications to tdBAG which make it consistent with the 
standard NJL approach: 
our vector interaction enhanced bag model (vBAG) accounts in a parameterized form for 
(a) the flavor dependent restoration of chiral symmetry, 
(b) modifications of the EoS due to vector interactions, 
and (c) - in the spirit of the tdBAG - a phenomenological correction to the EoS due to the deconfinement transition which one could otherwise not account for. 
Once parameterized, vBAG is as convenient to apply as the original tdBAG and therefore well suited for subsequent studies of dense and deconfined quark matter at finite densities and temperatures in astrophysics.

The manuscript is organized as follows:
In sec.~2 we review briefly the in-medium DS gap equations which we then reduce to the NJL model 
by assuming a simple contact interaction for the gluon propagator in sec.~3. 
In sec.~4 we begin to introduce our vBAG model and illustrate 
the small impact of the residual dynamical chiral symmetry breaking (D$\chi$SB)
on the EoS in the chirally nearly restored phase in the NJL model.
We also point to the importance of the flavor dependence of the chiral transition. 
In sec.~5 we add vector interactions to the model and obtain a set of equations which
distinguish our vBAG from the standard tdBAG. 
We then discuss a setup for a phase transition construction which
can ensure a simultaneous chiral and deconfinement phase transition in sec.~6,
and apply the resulting quark-hadron hybrid EoS to describe high mass compact stars in sec.~7. 
In sec.~8 we review the long standing hypothesis of absolutely stable strange 
quark matter as the ground state of matter, and finally conclude with a summary in sec.~9.

\section{Dyson Schwinger formalism with a contact interaction}

The key quantity of our analysis is the in-medium propagator of a single quark flavor \citep[cf.][]{Rusnak:1995ex,Roberts:2000aa},
\begin{equation}
\label{eq:QuarkPropagator}
S^{-1}(p;\mu) = i\vec \gamma \vec p A(p;\mu) + i\gamma_4\tilde p_4 C(p;\mu) + B(p;\mu)~,
\end{equation}
with $\tilde p_4=p_4+i\mu$, where $\mu$ is the chemical potential.  The gap solutions $A$, $B$, and $C$ are obtained from the gap equation
\begin{eqnarray}
\label{eq:QuarkPropagatorSigma}
S^{-1}(p;\mu) = i\vec \gamma \vec p+ i\gamma_4\tilde p_4+ m+\Sigma(p;\mu),
\end{eqnarray}
\begin{equation}
\label{eq:SelfEnergy}
\Sigma(p;\mu)=
\int_\Lambda\frac{d^4q}{(2\pi)^4}g^2D_{\rho\sigma}(p-q)\gamma_\rho\frac{\lambda^a}{2}S(q;\mu)\Gamma^a_\sigma(p;q),
\end{equation}
where $m$ is the current-quark mass, $D_{\rho\sigma}(p)$ the gluon propagator and
$\Gamma^a_\sigma(p;q)$ the quark-gluon vertex. 
$\Lambda$ represents a regularisation mass scale which, in a realistic treatment, would
be removed from the model by taking the limit $\Lambda\to\infty$. 
For the NJL model this procedure fails and $\Lambda$ usually represents a simple
cut-off for momentum integrals.
Following the approach from recent vacuum studies, 
cf. \citet{GutierrezGuerrero:2010md}, we apply the rainbow truncation 
$\Gamma^a_\sigma(p;q) = \frac{\lambda_a}{2}\gamma_\sigma$ 
and assume a contact interaction in momentum space, 
$g^2D_{\rho\sigma}(p-q)=\delta_{\rho\sigma}\frac{1}{m_G^2}\Theta(\Lambda^2-\vec p^2)$. 
The Heaviside function $\Theta$ provides a 3-momentum cutoff for all momenta $\vec p^2>\Lambda^2$ 
in order to regularize ultraviolet divergences inherent in Eq.~\eqref{eq:SelfEnergy}.
Different regularisation procedures are available\footnote{In fact, the regularisation scheme  
does not have to affect ultraviolet divergencies only. E.g., the very infrared, gluon mediated behavior is
ill described in NJL type models; an IR cutoff scheme can remove unphysical implications \citep{Ebert:1996vx}.}, we chose this hard cut-off scheme to
match our model with tdBAG, i.e. describe quarks as a quasi ideal gas of Fermions.
$m_G$ is a gluon mass scale which in this model simply defines the coupling strength. 
These approximations are sufficient to solve Eq.\eqref{eq:QuarkPropagatorSigma}. 
The $A$-gap has only a trivial, medium independent solution, $A=1$. 
The remaining gaps take the following form,
\begin{eqnarray}
\label{eq:BGap}
B_p &=& m + \frac{16N_c}{9m_G^2}\int_\Lambda\frac{d^4q}{(2\pi)^4}
\frac{ B_q}{\vec q^2A^2_q+ \widetilde q^2_4C^2_q + B^2_q}~,
\\
\label{eq:CGap}
\widetilde p_4^2 C_p &=& \widetilde p_4^2 + \frac{8N_c}{9m_G^2}\int_\Lambda\frac{d^4q}{(2\pi)^4}
\frac{\widetilde p_4\widetilde q_4C_q}{\vec q^2A^2_q+ \widetilde q^2_4C^2_q + B^2_q}~.
\end{eqnarray}
The integrals do not explicitly depend on the external momentum $p$. Consequently, both gap solutions are constant and vary only with $\mu$. One easily verifies from Eq.\eqref{eq:CGap} that the vector gap $C$ induces a medium dependent shift $\omega$ of the chemical potential $\mu$,
\begin{eqnarray}
\widetilde p_4 C = \widetilde p_4+ i\widetilde \omega = p_4 + i\mu^*\equiv\hat p_4~,
\end{eqnarray}
where the effective chemical potential $\mu^*$ is introduced in order to maintain a quasi particle description of the quarks. The vector condensate $\omega$ is given by
\begin{equation}
\label{eq:omega}
\omega=\mu^* - \mu = -\frac{8N_c}{9m_G^2}\int_\Lambda\frac{d^4q}{(2\pi)^4}\frac{i\hat q_4}{\vec q^2+ \hat q_4^2 + B^2}~.
\end{equation}
Instead of solving this equation to determine $\mu^*$ for a given $\mu$ it is more convenient to fix an arbitrary effective chemical potential $\mu^*$ and determine the actual chemical potential $\mu$ {\it post priori}. Then only the remaining Eq.~\eqref{eq:BGap}, rewritten in terms of the effective chemical potential $\mu^*$, needs to be solved self-consistently,
\begin{equation}
\label{EQ:DS_MassGap}
B = m +B \frac{16 N_c }{9m_G^2}\int_\Lambda\frac{d^4q}{(2\pi)^4}\frac{ 1}{\vec q^2+ \hat q^2 + B^2}~.
\end{equation}
In analogy to Eq.\eqref{eq:omega} the scalar condensate is defined as $\phi=B-m$. 
As we aim to provide an EoS at finite densities {\it and} temperatures we
apply Matsubara-summations to obtain the following set of equations,
\begin{eqnarray}
\label{EQ:massgapscalar}
B &=& m+\frac{4N_c}{9m_G^2}n_s(\mu^*,B)\\
\label{EQ:mugapvector}
\mu &=& \mu^*+\frac{2N_c}{9m_G^2}n_v(\mu^*,B),
\label{eq:gaps_DS}
\end{eqnarray}
with the single-flavor scalar and vector densities, $n_s$ and $n_v$, of an ideal spin-degenerate Fermi gas,
\begin{eqnarray}
n_s &=& 2\sum_\pm\int_\Lambda\frac{{\rm d}^3 \vec p}{(2\pi)^3}\frac{B}{E}
\left(
\frac{1}{2}-\frac{1}{1+\exp{(E^\pm/T)}}
\right), \\
n_v &=& 2\sum_\pm\int_\Lambda\frac{{\rm d}^3 \vec p}{(2\pi)^3}\frac{\mp1}{1+\exp{(E^\pm/T)}},
\end{eqnarray}
with $E^2=\vec p^2+B^2$ and $E^\pm=E\pm\mu^*$.

\begin{figure*}[htp!]
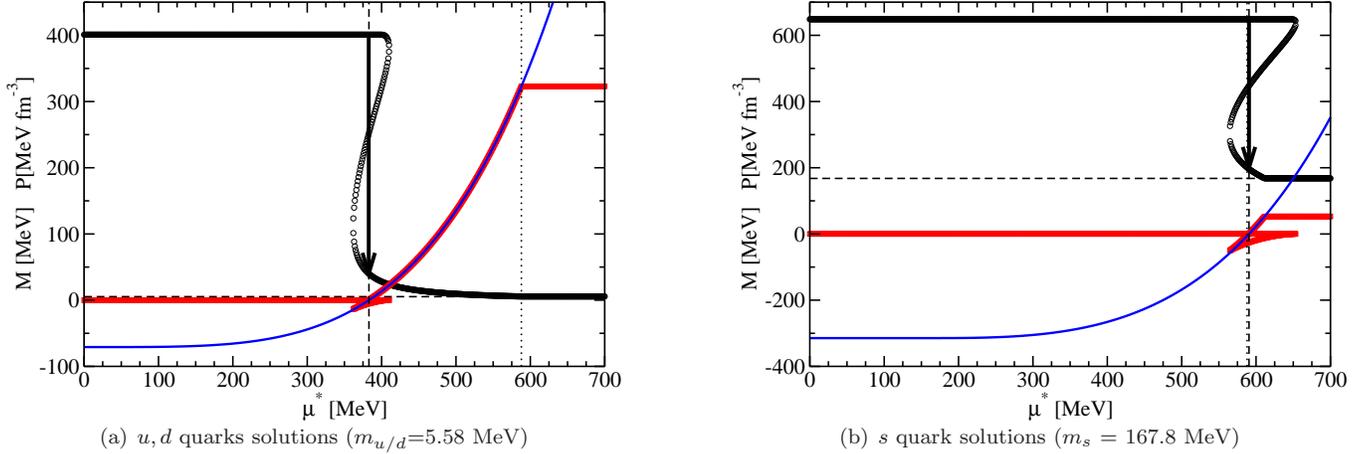

\subfigure[$u,d$ quarks solutions ($m_{u/d}$=5.58 MeV)]{
\includegraphics[width=0.95\columnwidth]{MP_mustar_lightquark}
}
\hfill
\subfigure[ $s$ quark solutions ($m_s$ = 167.8 MeV)]{
\includegraphics[width=0.95\columnwidth]{MP_mustar_heavyquark}
}
\caption{
\label{FIG:MassGapPress}
(color online) Single flavor dynamical masses (black) and corresponding pressure (red) computed within the NJL model. The latter is well fitted by the pressure of an ideal Fermi gas (with bare quark mass $m_f$) shifted by a chiral bag constant $B_\chi$ (blue). Parameters: set~IV, table \ref{TAB:NJL}.}
\end{figure*}

It is common for DS calculations to determine the thermodynamic pressure in steepest descent approximation. It consists of ideal Fermi-gas and interaction contributions,
\begin{eqnarray}
P_{FG}&=&{\rm Tr}\ln S^{-1}\nonumber\\ &=& 2N_c\int_\Lambda \frac{d^4 p}{(2\pi)^4}\ln(\vec p^2 + \hat p_4^2 + B_\mu^2)~, \\
P_I &=& -\frac{1}{2}{\rm Tr}\Sigma S=\frac{3}{4}m_G^2\omega^2-\frac{3}{8}m_G^2\phi^2~.
\end{eqnarray}
The pressure of the free Fermi gas consists of vacuum and kinetic parts, $P_{FG}=P^{vac}_{FG}+P^{kin}_{FG}$, with
\begin{eqnarray}
P_{FG}^{vac}&=&
 2N_c\int_\Lambda\frac{{\rm d}^3 \vec p}{(2\pi)^3}
E,
\\
P_{FG}^{kin}&=&2TN_c\sum_\pm\int_\Lambda\frac{{\rm d}^3 \vec p}{(2\pi)^3}
\ln
\left(
1+\exp\left(-\frac{E^\pm}{T}\right)
\right).
\;\;\;\;\;\;
\end{eqnarray}
Divergences occur only in the vacuum contributions to the total pressure, given as
\begin{eqnarray}
\label{EQ:PVac}
P_{vac}&=& 2N_c\int_\Lambda\frac{{\rm d}^3 \vec p}{(2\pi)^3}
E
\nonumber\\
&&+
\frac{8N_c^2}{9m_G^2}
\left(
\int_\Lambda\frac{{\rm d}^3 \vec p}{(2\pi)^3}
\frac{B_0}{E}
\right)^2,
\end{eqnarray}
where the 2$^{nd}$ term on the right-hand side originates from the scalar condensate $\phi$.
$B_0$ is the dressed quark mass at $\mu=T=0$. 
The remaining medium dependent pressure is free of divergences and does, 
in a technical sense, not require regularisation.

\section{NJL model comparison}

We compare the previous results to a simple model of the NJL-type. The Lagrangian is separated into free and interaction parts, $\mathcal{L}_0$ +$\mathcal{L}_{\rm I}$, where the latter accounts for scalar and vector interactions,
\begin{equation}
\label{EQ:LInteract}
\mathcal{L}_{\rm I}=\mathcal{L}_{\rm S}+\mathcal{L}_{\rm V} = G_s\sum_{a=0}^8(\bar q_f\tau_a q_f)^2 + G_v(\bar q_f i\gamma_0 q_f)^2~,
\end{equation}
with single-flavor quark and anti-quark spinors $(q_f,\bar q_f)$ and without flavor-mixing terms. 
In the following paragraphs, we will not elaborate further on NJL type models nor on 
the techniques to proceed from the Lagrangian to describe the thermodynamics of the system. 
These are well described in the literature and for this particular model explained in detail, e.g., in \cite{Buballa:2003qv}. 

The thermodynamic pressure of a single flavor $f$ is obtained from the grand canonical thermodynamic potential, 
\begin{eqnarray}
\label{EQ:NJLpotential}
P=-\Omega_f =P_{FG}- \frac{\tilde\phi^2}{4G_s} + \frac{\tilde\omega^2}{4G_v}~,
\end{eqnarray}
with the scalar and vector condensates, $\tilde\phi$ and $\tilde\omega$. 
Again, the free part describes quarks as ideal quasi-particles with effective masses, $m^*=m+\tilde\phi$, and effective chemical potentials, $\mu^*=\mu-\tilde\omega$. 
The corresponding NJL model gap equations are
\begin{eqnarray}
\tilde\phi=2 G_s N_c n_s(\mu_f^*,m_f^*),\tilde\omega = 2 G_v N_c n_v(\mu_f^*,m_f^*)~.
\label{eq:gaps_NJL}
\end{eqnarray}
Eqs.~\eqref{eq:gaps_NJL} exactly reproduce the gap Eqs.~\eqref{eq:gaps_DS} derived within the DS framework, if one identifies
\begin{equation}
\frac{1}{2}G_s=G_v=\left(\frac{1}{3m_G}\right)^2.
\end{equation}
The relation of the NJL model coupling constants, $G_v=G_s/2$, is consistent with the result obtained after Fierz transformation of the one-gluon exchange interaction \citep[cf.][]{Buballa:2003qv}. We summarize the first part of this work by observing that the results we obtained within the DS approach under the assumption of a contact interaction in the gluon sector describe the same thermodynamics as the briefly reviewed NJL model. 
We conclude that in the limiting case 
of a contact interaction both approaches, DS and NJL, give identical results.
Table \ref{TAB:NJL} contains specific NJL model parameterisations taken from \citet{Grigorian:2006qe}, reproducing pion and kaon masses, and the pion decay constant for different light quark constituent masses $M_0^{u/d}$.
For the purpose of our analysis we restrict ourselves to set I and IV with the smallest and largest constituent 
light quark masses in vacuum.

\section{chiral bag model}

In this and the following section we introduce a modified bag model which can be parameterized to reproduce NJL model results in very good agreement. We extend it afterwards to additionally mimic confinement.

Fig.~\ref{FIG:MassGapPress} shows the dynamical quark masses of $u,d,$ (left) and $s$ (right) quarks 
and the corresponding pressure. 
Chiral symmetry is broken at small and restored at large  $\mu$. 
The intermediate region is characterized by the existence of multiple mass gap solutions. 
The thermodynamically stable solution maximizes the pressure of the system at given $\mu$. 
As indicated by the arrow this results in a sharp transition from the branch of chirally broken ($\chi$B) to the branch of chirally restored  ($\chi$R) solutions. 
The multiple solutions of our gap equations are a feature not restricted to the NJL model.
It owes to the nonlinearity of QCD's gap equations in general \citep[cf.][]{Chang:2006bm,Wang:2012me,Raya:2013ina}.
We now fit the pressure of the $\chi$R branch by calculating only the kinetic pressure of an ideal gas with the bare quark mass $m_f$ and shift it by a constant but flavor dependent value $B^f_\chi$. 
This procedure can be understood as a complete transfer of the scalar, regularisation dependent vacuum contributions to the pressure according to Eq.\eqref{EQ:PVac} into $B^f_\chi$. 
That NJL models in the $\chi$R domain can be fitted well by this approach has been shown for  a flavor blind effective bag constant by \citet{Buballa:1998pr}, and later for flavor dependent bag constants in \citet{Buballa:2003qv}.

In analogy to the bag pressure, we name  $B^f_\chi$ the chiral bag pressure. 
Evidently, this fit gives good results in the $\chi$R domain we are interested in. 
The value of $B^f_\chi$ is obtained from the condition $ B^f_\chi =P(M_f^0,\mu_f=0)-P(m_f,\mu_f=0)$, where $M_f^0$ is the dressed vacuum quark mass \citep[cf.][]{Buballa:2003qv}. 
This prescription has first been applied in a DS framework to determine the bag constant by \citet{Cahill:1985mh}.
$B^f_\chi$ depends on vacuum properties only. The critical chemical potential $\mu_\chi^f$ at which chiral symmetry is restored in this approximation is defined by the relation $P_{kin}^f(m_f,\mu_\chi^f)= B^f_\chi$. 
The model fits $\chi$R quark matter according to this prescription for $\mu^f\ge\mu_\chi^f$, below $\mu_\chi^f$ the  pressure of the single quark flavor and all related quantities can be set to zero as this sector is characterized by confined hadrons, consequently the pressure and density of deconfined quarks is zero. 
The existence of a flavor dependent gap $B^f_\chi$ is a feature not inherent to tdBAG. As we illustrated, it originates explicitly from the breaking of chiral symmetry -- an effect which has been deliberately ignored in tdBAG as stated by \cite{Farhi:1984qu}. 
Correctly,  \cite{Farhi:1984qu} refer to the tdBAG bag constant as a quantity that mimics confinement and pointed out clearly, that the model does not account for chiral symmetry breaking. Later in this paper we will point out, how these different bag constants are related. The values of $\mu_\chi^f$ and corresponding chiral 
bag constants $B^f_\chi$ for different NJL model parameters are listed in table~\ref{TAB:CBAG}.

\begin{table}[htp]
  \caption{NJL parameterisations according to \cite{Grigorian:2006qe}.}
  \label{TAB:NJL}
  \begin{center}
  \leavevmode
    \begin{tabular}{c|cccccc} \hline \hline              
  	        & $M^{u/d}_{0}$	&   $M^s_0$	& $G_S\Lambda^2/2$   	& $\Lambda$    	& $m_{u/d}$ 	&  $m_s$		\\ 
   		& $[$MeV$]$ 		& $[$MeV$]$ 	& 					&   $[$MeV$]$	& $[$MeV$]$ 	&  $[$MeV$]$ 	\\ \hline 
  I 		& 330.0			& 610.0		& 2.17576				& 629.540		& 5.27697		& 171.210		\\
  II 		& 367.5			& 629.6		& 2.31825				& 602.472		& 5.49540		& 170.417		\\
  III 		& 380.0			& 636.5		& 2.36582				& 596.112		& 5.54297		& 169.559		\\
  IV		& 400.0 			& 647.7		& 2.44178				& 587.922		& 5.58218		& 167.771		\\\hline\hline
    \end{tabular}
  \end{center}
\end{table}

\section{vector interaction in the bag model}

Before we address this issue in detail we extend the chiral bag model to account for the repulsive vector interaction.
This aspect is crucial if one aims to describe stable and massive hybrid neutron star configurations with quark matter cores, cf.~\citet{Klahn:2013kga}, in agreement with the recent observations of pulsars with two solar masses by \citet{Demorest:2010} (PSR J1614-2230 with $1.97\pm 0.04$ M$_\odot$) and \citet{Antoniadis:2013} (PSR J0348-0432 with $2.01\pm 0.04$ M$_\odot$). 

\begin{figure}[htp!]
\centering
\includegraphics[width=0.9\columnwidth]{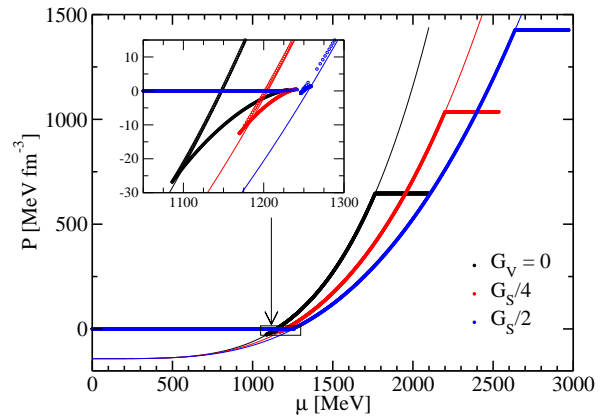}
\caption{
\label{FIG:VectorCoupling}
(color online) Impact of vector interactions on the stiffness of the EoS and the corresponding shift of the critical potential $\mu_\chi$ which defines the chiral phase transition. Thin lines: vBag.  Bold lines: NJL model.  Parameters: set IV, table~\ref{TAB:NJL}.}
\end{figure}

The assumption that the dressed quark mass in the $\chi$R region is effectively equal to the bare quark mass $m$ and the pressure of the system evolves mainly due to the corresponding kinetic contribution while all effects of the scalar condensate are hidden in the constant offset $B_\chi^f$ frees us from solving the mass gap Eq.~\eqref{EQ:massgapscalar}, the only equation which had to be solved self consistently. The actual chemical potential $\mu$ is easily obtained from Eq.~\eqref{EQ:mugapvector} which ensures that the vector gap Eq.~\eqref{eq:CGap} is solved correctly.  It requires to perform only a single integration in order to compute 
the one-particle number density $n_f(\mu_f^*)\equiv n_{v}(\mu_f^*,m_f)$. All further calculations are simple rescaling operations involving free Fermi-gas expressions in terms of the effective chemical potential $\mu^*$. The rescaling results from the vector-gap equation Eq.\eqref{eq:gaps_NJL} and affects the thermodynamics in terms of pressure and energy density, 
\begin{eqnarray}
\label{eq:muscale}
&& \mu_f=\mu_f^*+K_v n_f(\mu_f^*)~, \\ 
&& P_f(\mu_f)=P_{FG,f}^{kin}(\mu_f^*)+\frac{K_v}{2} n_{f}^2(\mu_f^*) - B^f_\chi~,
\label{eq:vbag_p}
\\
&& \varepsilon_f(\mu_f)=\varepsilon_{FG,f}^{kin}(\mu_f^*)+\frac{K_v}{2} n_f^2(\mu_f^*)+B^f_\chi~,
\label{eq:vbag_e}
\\
&&
n_f(\mu_f)=n_f(\mu_f^*).
\end{eqnarray}
These equations (except for a few remaining considerations regarding deconfinement in sec.6) 
define the novel vBAG model and are sufficient to reproduce NJL model results with finite values for the vector coupling. For convenience we introduced a new coupling constant $K_v=2 G_v$. In Fig.~\ref{FIG:VectorCoupling} we reproduce NJL model results for different values of the vector coupling constant $G_V$ at a fixed value for the scalar coupling constant $G_S$ and compare them to the bag model approximation. As the scalar interaction is unchanged the vacuum offset $B^f_\chi$ is the same for all shown scenarios. The critical chemical potential $\mu_\chi^f$ shifts to higher values because of the increasing vector coupling which increases the chemical potential according to Eq.~\eqref{eq:muscale}. This effect dominates in comparison to the additional contribution to the pressure (2$^{nd}$ term in Eq.~\eqref{eq:vbag_p}) which just by itself would increase $P$ at given $\mu^*$ and therefore shift $\mu_\chi^f$ towards smaller values.

\section{confinement transition}

\begin{figure}[tp!]
\centering
\includegraphics[width=0.9\columnwidth]{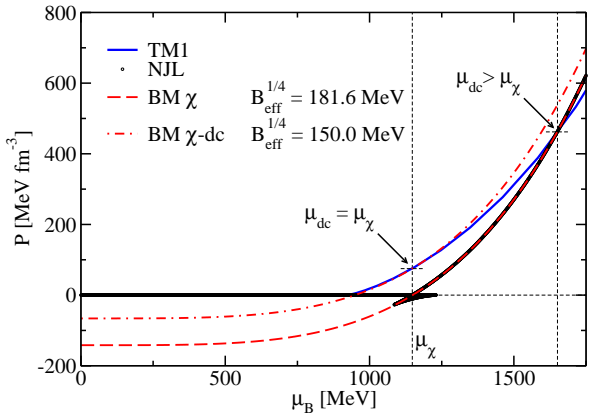}
\caption{
\label{FIG:PhaseTransitionSymMatter}
(color online) Symmetric matter phase transition from nuclear to (u,d)-flavor quark matter using the nuclear model EoS TM1(blue line), from \citet{Shen:1998gg}, NJL model for 2-flavor quark matter using the same parameterization as in Fig.~\ref{FIG:MassGapPress} (black circles) and the bag model parameterization of the NJL result (red lines). Applying the standard Maxwell construction the crossings with TM1 are thought to mimic the deconfinement phase transition. Red dashed: By default, the so defined deconfinement chemical potential $\mu_{dc}$ is larger than $\mu_\chi$. Red dash-dotted: The pressure is shifted by a positive value $P_{dc}=P_{{\rm TM1}}(\mu_\chi)$ to match $\mu_{dc}=\mu_\chi$ when the pressure of quark and nuclear matter equal. This models simultaneous deconfinement and chiral symmetry restoration. Parameters: set IV, table~\ref{TAB:NJL}.}
\end{figure}

The model as introduced so far is suited for deconfined quarks but as typical for NJL models it
cannot describe truly confined matter\footnote{However, confinement understood as the absence of quark production
thresholds can be mimicked even in the non renormalizable NJL model \citep[cf.][]{GutierrezGuerrero:2010md}.}. 
Already accounting for bound states is a demanding problem that requires analyses 
of two- and three-particle in-medium correlations. 
Only few studies exist which aim to describe dense hadronic matter explicitly in terms of quark matter degrees of freedom, 
e.g., by \citet{Wang:2010iu} and \citet{Blaschke:2013zaa}. 
For phenomenological purposes deconfinement is therefore usually modeled via a first order Maxwell transition from a hadronic to a quark matter EoS which both are not modeled within the same framework. 
The critical chemical potential $\mu_{dc}$ which defines the deconfinement transition is obtained by applying
the Maxwell-condition $P_{H}(\mu_{dc})=P_{Q}(\mu_{dc})$. We illustrate this in Fig.~\ref{FIG:PhaseTransitionSymMatter} for isospin-symmetric matter {\it at zero temperature}, where the nuclear EoS is the relativistic mean-field model TM1 of \citet{Sugahara:1993wz} and \citet{Shen:1998gg}. 
A TM1 based EoS by \citet{Hempel:2009mc} can be found online in the CompOSE-database \footnote{http://compose.obspm.fr/spip.php?article29}. CompOSE is explained in detail in \citet{Typel:2013rza}. 
The following discussion does not rely on TM1 as a specific choice; our arguments hold in general
and the same qualitative results would be obtained for any different nuclear EoS.

For the quark matter branch we neglect the strange flavor and consider a purely chiral bag model for iso-spin symmetric two-flavor quark matter ($n_u=n_d$). With the chosen parameterization we obtain the two-flavor chiral bag constant $B_\chi^{u,d}=B_\chi^{u}+B_\chi^{d}=(181.6 $ MeV$)^4$, which results in a chiral critical chemical potential of $\mu_{\chi}=1148$ MeV. The Maxwell-condition without further modifications to the model is then fulfilled at a deconfinement critical chemical potential of $\mu_{dc}=1651$ MeV. 
The intermediate region spans an interval of about 500 MeV in which the physical interpretation is ambiguous.
 Our main concern is that the QM model predicts the restoration of chiral symmetry which should be reflected by the nuclear model EoS which is of course not the case, as TM1 is an independent model. There are several possibilities to deal with this situation. The first would be to accept it as a poor description of hypothetic quarkyonic matter where the chiral symmetry of light quarks is restored in a gas of still confined nucleons \citep[cf.][and references therein]{McLerran:2007qj,Glozman:2008ja}. Microscopic studies indicate, that the transitions related to chiral restoration and deconfinement are closely related and can occur at the same or very similar chemical potential \citep[cf.][]{Bender:1996bm, Bender:1997jf,Blaschke:1997bj,Qin:2010nq}. In order to illustrate the setup of our fully defined model we prefer such a situation and assume that chiral restoration and deconfinement occur at the same critical chemical potential, hence $\mu_{\chi}=\mu_{dc}$. We model a hybrid EoS which fulfills this criterion and introduce the additional {\em deconfinement} bag constant, $B_\text{dc}$, which is added to the overall pressure of our quark matter model such that $P_{H}(\mu_{{\chi/dc}})=:P_{Q}(\mu_{\chi/dc})$. The same strategy has been suggested in \citet{Pagliara:2007ph} as a sound alternative to just perform a standard Maxwell phase transition from a nuclear model EoS to a vacuum renormalized NJL quark matter EoS. While we consider this as a likely scenario we point out, that a rather quarkyonic behavior with $\mu_\chi \ne \mu_{dc}$ is not ruled out. 
 A strength of our model is the independent treatment of the deconfinement transition and chiral restoration, where the latter is defined only by the chiral bag constants and the relation $P_{kin}^f(m_f,\mu_\chi^f)= B^f_\chi$, whereas the deconfinement critical potential depends on all chiral bag constants $B^f_\chi$ {\it and} the nuclear model EoS. The total pressure in our model is $P=\sum_f (P_f^{kin} + K_v n_f^2/2) - B_{eff}$ with 
 \begin{equation}
 B_{eff}=\sum_f B_\chi^f - B_\text{dc}.
 \end{equation}
We emphasize, that a single quark flavor can contribute to the sum of chiral bag constants only if chiral symmetry for this
flavor is restored; as explained in sec.~4. 
This is of particular importance for the $s$-quark if $\mu_\chi^s>\mu_\chi^{u,d}$.
Note, that this prescription -- subtracting an effective bag constant from the sum of all partial kinetic pressures -- corresponds to the standard tdBAG  prescription. 
 In addition to tdBAG, our model identifies different contributions to the effective bag constant. This is not a minor technical detail. One has to keep in mind, that the original tdBAG explicitly omits the scalar (and vector) interaction and attributes the bag constant fully to a confining background field with positive energy and hence negative, confining pressure. In our prescription, the positive value of the bag constant results {\it only} from the restoration of chiral symmetry, while confinement/deconfinement, although introduced merely phenomenologically, reduces this value. From our perspective, this makes perfect sense if one recalls, that the most naive perception of confinement is the binding of quarks in the chiral broken phase. Thus, our deconfinement bag constant is easily interpreted as the actual binding energy of confined quarks which effectively reduces the free energy of the system.

\section{Neutron stars with QM core}

In this section we discuss how vBAG clears the long lasting perception originating from tdBAG, that quark matter tends to be too soft towards higher densities in order to support the idea of QM in compact stars.

\begin{figure}[ht]
\centering
\includegraphics[width=\columnwidth]{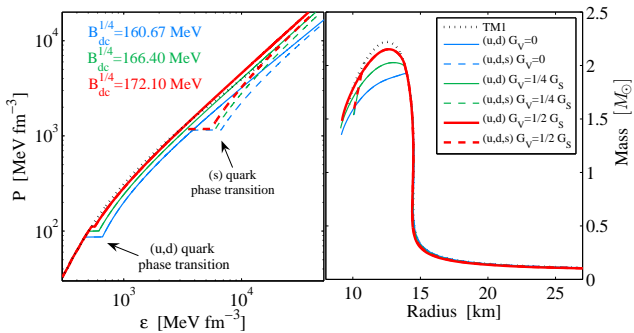}
\caption{
\label{FIG:eos_mr}
(color online) vBAG EoS pressure vs. energy density for neutron star matter (left panel) and corresponding mass-radius relations (right panel) for $B_{(u,d),\chi}^{1/4}=137.6$~MeV and $B_{s,\chi}^{1/4}=224.1$~MeV (set I, table~\ref{TAB:CBAG}), comparing different values of $G_V$. Values of the deconfinement bag constants for the different $G_V$ are listed in the left panel.
NJL parameters: set I, table \ref{TAB:NJL}.}
\end{figure}

In the left panel of Fig.~\ref{FIG:eos_mr} we illustrate the phase transition in $\beta$-equilibrated neutron star matter for selected chiral bag constants, $B_{(u,d),\chi}^{1/4}=137.6$~MeV and $B_{s,\chi}^{1/4}=224.1$~MeV, and varying vector interaction strengths ($G_v = (0,1/4,1/2)G_s$). 
$B_{dc}$ has been adjusted such that $\mu_\chi=\mu_{dc}$. $\mu_\chi$ varies for different values of the vector coupling $G_v$,  as illustrated by the inset of Fig.\ref{FIG:VectorCoupling}, and therefore the value of $B_{dc}$ depends on $G_v$ as well. Variations of the $s$-quark onset density are almost negligible. 

Due to the large vacuum mass of the $s$-quark, the phase transition from TM1 to two-flavor ($u,d$) quark matter 
takes place at lower density, followed by the transition to three-flavor ($u,d,s$) matter at higher density. 
This is the behavior one expects from NJL-type models without flavor coupling channels. 
It is not accounted for by  tdBAG which ignores D$\chi$SB and consequently predicts a transition from nuclear to three-flavored strange matter.
In contrast, vBag describes a sequential transition from nuclear to two-flavor, then to three-flavor quark matter,
where the three flavor branch of the EoS is found only for unstable NS configurations, 
see the right panel of Fig.\ref{FIG:eos_mr}.

Note that for the canonical vector coupling of $G_V=1/2~G_S$ vBAG nearly reproduces the nuclear model EoS TM1. 
This ``masquerade'' has been discussed by \citet{Alford:2004pf}.
It results in a very similar mass-radius relation for pure neutron and hybrid stars for this parameter set (right panel of Fig.~\ref{FIG:eos_mr}) . 
The different onset-densities for two-flavor matter at different values of $G_v$ are reflected in different neutron star mass-radius relations. For $G_v=0$ the transition in general results in unstable configurations for the parameters explored here. Hence, the vector interaction is essential to stabilize the compact stellar object. Larger values of $B^{u,d}_\chi$ associated with larger quark masses result in higher critical densities for the phase transition but qualitatively reproduce the above discussed features as long as the transition density does not reach values where already the purely nuclear NS configurations render unstable.

\section{stability analysis of strange matter}

The long-standing hypothesis of absolutely stable strange quark matter as the ground state of strongly interacting matter, introduced by \citet{Witten:1984}, is supported by the standard tdBAG of \citet{Farhi:1984qu}. As stated before, tdBAG does not account for D$\chi$SB nor the vector interaction channel. 
NJL model studies imply that taking D$\chi$SB into account absolutely stable strange matter can be ruled out, see e.g. \citet{Buballa:1998pr}. 
Our model is a phenomenological hybrid of NJL and tdBAG and, in terms of the defining parameters, namely $B^f_\chi$, $B_{dc}$ and $K_v$, could in principle be used to argue against or in favor of absolutely stable strange matter 
if one decides to choose the corresponding values freely. 

\begin{figure}[h]
\centering
\includegraphics[width=0.9\columnwidth]{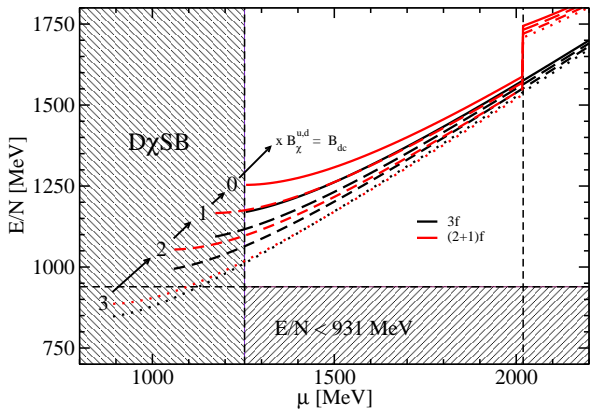}
\caption{
\label{FIG:strange}
(color online) $\beta$-equilibrated strange quark matter: energy for different values of $B_{dc}$ at $T=0$, compared with the nuclear matter minimum (horizontal shaded band). The chiral bag constant $B_\chi$ has been chosen from the parameter set I, table~\ref{TAB:NJL}
with a dressed light quark mass (in vacuum) of 330 MeV to provide a number close to the lower limit
of $B_\chi$. 
red: finite $B_\chi^s$, hence there are two {\it sequential} chiral transitions (nucl.$\leftrightarrow$2f$\leftrightarrow$3f).
black: $B_\chi^s\equiv 0$ results in only one chiral transition (nucl.$\leftrightarrow$3f).
Note that the vertical shaded band marks the domain of D$\chi$SB which tdBAG {\it cannot describe}.
See text for more details.
}
\end{figure}

Briefly described, the concept of absolutely stable strange matter is based on
the idea, that two flavor matter at zero pressure has a larger energy per particle than the most stable nuclei (we chose a value of 931 MeV) because otherwise these would decay into the constituent $u$- and $d$- quarks. 
Adding $s$-quarks to the mixture, however, lowers the energy per particle. Hence strange matter could be more stable than nuclear matter \citep[cf.][]{Farhi:1984qu}. This leaves a small window of possible (small) bag constants which support this idea. 
The main reason why chiral models do not confirm this idea is, that the chiral bag constant itself is flavor dependent due to D$\chi$SB and does not differ from the 2-flavor bag constant due to the large value of the dressed $s$-quark mass and the resulting high densities where free $s$-quarks can appear. 
More precisely, tdBAG predicts absolutely stable strange matter in a domain 
where chiral models predict the chiral symmetry for the $s$-quark to be broken
and to be restored for the light quark.
This is the definition of two flavor quark matter where $s$-quarks do not exist. 
This situation holds in the vertical gray shaded band (D$\chi$SB) in Fig.\ref{FIG:strange}.
In the domain where the $\chi$-symmetry of the $s$-quark is restored
the energy per particle of the 2-flavor phase is already high and would be increased further by the chiral phase transition. This is illustrated by the red solid line in Fig.~\ref{FIG:strange} with $B_\chi^s>B_\chi^{u,d}$
where $\mu_\chi^s$ is indicated by the black vertical dashed line near 2~GeV.

In general, the chiral bag constant as obtained from NJL model analyses has a large value of $B_\chi^{1/4}\ge 160$ MeV already in the two flavor case, a value which exceeds the one predicted for stable strange matter. 
Therefore, we use the additional free parameter $B_{dc}$ to reduce the effective bag constant 
and the energy per particle. Here we varied $B_{dc}$ = (0,1,2,3) times $B_{\chi}^{u/d}$.
In order to reproduce the assumptions of tdBAG we set $B_\chi^{(s)}=0$ (black lines in Fig.~\ref{FIG:strange}). It lowers the energy, however, only significantly at $\mu < \mu_\chi^{u,d}$, hence a region governed by D$\chi$SB of the light flavors where matter is still confined and a model for deconfined matter does not apply.
Note that the standard tdBAG would still predict a scenario with stable strange matter as it 
does not account for D$\chi$SB.  vBAG predicts effectively the same bag constant but excludes the existence of free quarks from this region, since the chemical potentials are below the 
critical chemical potential for D$\chi$SB, $\mu<\mu_\chi$.

The last available parameter is $K_v$. From Eq.~\eqref{eq:vbag_e} follows, that the vector interaction adds to the energy density
and therefore makes strange matter less stable at given density. Moreover, as discussed earlier, it shifts the chiral transition to higher chemical potentials which further disfavors stable strange matter. 

Back to our analysis, we conclude that the idea of stable strange matter is not supported
for a NJL based parameterization of the chiral bag constants,
even after introducing a deconfinement bag constant, which leads to effective bag constants
similar to the original tdBAG.

How general is this result?
The DS study of  \citet{Nickel:2006kc} implies that the color flavor locked phase (where all three quark flavors exist) is favored over the 2SC phase which consists of the two light quarks only.
Would this make absolutely stable strange matter more likely? We claim that not.
The crucial question regarding the stability of strange matter is the onset of chiral symmetry
restoration in terms of the chemical potential for all three flavors.
This is, because  absolutely stable strange matter is predicted in tdBAG as a chiral symmetric phase.
As soon as the light quarks attain mass due to D$\chi$SB the corresponding baryon energy per particle
has to be either higher (no confinement) or equal (confinement) to that of nuclear matter.
If it would be lower, the argumentation of \citet{Farhi:1984qu} holds: nucleons would
simply decay into their constituent light quarks.
Our simple NJL model predicts D$\chi$SB for the light quarks at critical $\mu_\chi$ beyond values
which support the absolutely stable strange matter hypothesis.
In contrast, \citet{Nickel:2006kc} showed that the current mass of the $s$-quark and therefore $\mu_\chi^s$ reduce
due to color-flavor locking.
However, the color-flavor locking at the same time {\it increases} $\mu_\chi^{u,d}$,
in fact they coincide, $\mu_\chi^s=\mu_\chi^{u,d}$.
We argue, that this increase of the light quark critical chemical potential for D$\chi$SB
acts even stronger against the formation of absolutely stable strange matter.

We point out again, that the crucial mechanism which prevents the existence of 
absolutely stable strange matter is D$\chi$SB in the light quark sector, 
an explicitely feature inherent to non-perturbative QCD.
Therefore, any analyses of this hypothesis that bases on a perturbative approach to QCD, e.g., \citet{Kurkela:2009gj},
is not suited to derive reliable qualitative conclusions.

\section{summary}

Quark matter in compact stellar objects, such as (proto)neutron stars, has been studied widely based on the simple and powerful tdBAG model. 
With recent advances in mass-determinations of massive compact objects it has become increasingly difficult to fulfill the $\sim 2$~M$_\odot$ maximum mass constraints with tdBAG quark matter EoS. 
It becomes particularly difficult if the hadron-quark phase transition yields large latent heat, i.e. large jump between hadron and quark EoS.  In this article  we define the novel vBAG model by introducing vector interaction.
Since vector interactions stiffen the EoS with increasing density this provides an undisputed cure to the aforementioned problem without raising the technical difficulties in computing the EoS.

We derived vBAG from the currently most sophisticated approaches to dense quark matter. 
We demonstrate how a simple contact interaction on the gluon propagator level allows us to reproduce NJL and Bag models within the DS formalism in medium (i.e. at finite chemical potential). 
Moreover, we illustrated how the vector interaction channel is a crucial input for the investigation of hybrid, quark matter -- neutron stars. 
Confinement is mimicked by a confinement bag constant $B_{dc}$. 
It reduces the sum of all flavored chiral bag constants. 
Together, they reproduce effective bag constants typical for tdBAG~\citep[cf.][]{Farhi:1984qu}. 
We revisited the absolutely stable strange matter hypothesis and conclude that it cannot be easily discussed without a thorough understanding of D$\chi$SB in particular of the light quarks. 
Based on our extended bag model, which implements D$\chi$SB in a coarse fashion 
we confirm previous NJL model studies that rule out the existence of stable strange matter
and that tdBAG's prediction of stable strange matter results directly from the suppression of D$\chi$SB.
Confinement in terms of $B_{dc}$ is not likely to be an effect which would change this result.

The model EoS vBAG has been deliberately developed to be simple and easy applicable while catching the main (minimal) features we expect from QCD in dense matter, namely D$\chi$SB and to some extent deconfinement. 
We consider it an important improvement of the standard bag model as it tries to clearly separate these effects. 
Although the number of parameters is higher, each of them has a clear physical interpretation. 
Note, that QCD in medium is not plainly solvable, which was reminiscent when we motivated the model from the DS perspective.
We think that vBAG is a practical tool for modelers who wish to account for QCD degrees of freedom in complex dense systems -- in particular for applications in astrophysics studies.

\begin{table*}[htp!]
  \begin{center}
  \leavevmode
  \caption{Single flavor, effective two-flavor chiral bag constants and $\mu_{\chi/dc}$ for the parameterisations of table \ref{TAB:NJL}.}
  \label{TAB:CBAG}
        \begin{tabular}{c|cccc|cc|cccc} \hline \hline              
 	& \multicolumn{4}{c|}{chiral bag model parameters}		& && \multicolumn{4}{c}{Phase Transition TM1$\to$ 2f QM (symmetric)} 			\\\hline
	& ${P^{u}_{BAG}}^\frac{1}{4}$   	& $(\sum\limits_{u,d}P^i_{BAG})^\frac{1}{4}$ 	& ${P^s_{BAG}}^\frac{1}{4}$	& $(\sum\limits_{u,d,s}P^i_{BAG})^\frac{1}{4}$ & $\mu_\chi^{u/d}$ 	& $\mu_\chi^s$  	& $\mu_\chi$	& $\mu_{\rm dc}$ (Maxwell)	& $P_{\rm TM1}^{\frac{1}{4}}(\mu_\chi)$	& $B_{\rm eff}^\frac{1}{4}$ ($\chi$-dc) \\ 
 	& $[$MeV$]$ 					& $[$MeV$]$ 		&  $[$MeV$]$				& $[$MeV$]$ 							& $[$MeV$]$		& $[$MeV$]$		& $[$MeV$]$	& $[$MeV$]$				& $[$MeV$]$			& $[$MeV$]$\\ \hline 
I 	&	137.6						&  163.6								& 224.1					&  238.5								&	343			&	594			& 1029		&	1458					&	120.7	&  149.8		\\
II	&	145.8					&  173.4								& 221.9					&  240.2								&	365			&	591			& 1094		&	1569					&	141.4	&  149.8		\\
III	&	148.5					&  176.6								& 221.7					&  241.3								&	371			&	590			& 1114		&	1600					&	147.1		&  149.9		\\
IV	&	152.7					&  181.6								& 221.7					&  243.3								&	383			&	590			& 1148		&	1651					&	155.3	&  150.0            \\\hline\hline
    \end{tabular}
  \end{center}
\end{table*}

\section*{acknowledgements}
We thank F.~Weber, C.D.~Roberts, K.~Redlich, 
A.~Raya, G.~Pagliara, J.M.~Lattimer, M.~Hempel, and D.~Blaschke 
for their helpful comments and discussions.
Both authors are grateful for support by the Polish National Science Center (NCN) 
under grant number UMO-2013/09/B/ST2/01560 (T.K.)
and under grant number UMO-2013/11/D/ST2/02645 (T.F.),
and appreciate the support for networking activities 
provided by the COST Action MP1304 "NewCompStar".

%\bibliography{references}

\end{document}